# Measurement of secondary cosmic-ray neutrons near the geomagnetic North Pole

Richard S. Woolf[a], Laurel E. Sinclair[b], Reid A. Van Brabant[b], Bradley J. A. Harvey[b], Bernard F. Phlips[a], Anthony L. Hutcheson[a], Emily G. Jackson[c]

[a]Space Science Division, U. S. Naval Research Laboratory, 4555 Overlook Ave., SW, Washington, DC 20375

[b]Natural Resources Canada, Government of Canada, 930 Carling Ave (CEF, Bldg 7, Observatory Cr.) Ottawa, Ontario, K1A 0Y3, Canada

[c]Formerly: NRC Research Associate resident at the U. S. Naval Research Laboratory, 4555 Overlook Ave., SW, Washington, DC 20375, Now: Remote Sensing Laboratory, Joint Base Andrews, MD 20762

1. Introduction

Cosmic rays, either galactic or solar, are energetic charged particles from space that consist of protons (~90%), alpha particles (~9%) and other heavier nuclei (~1%). The Earth's magnetic field affects the minimum momentum per unit charge (magnetic rigidity) that a cosmic-ray particle incident at a given angle can have and still reach a given location above the Earth (Goldhagen et al., 2002). For vertically incident particles, this quantity is called the geomagnetic vertical cutoff rigidity, $R_C$, for that point; it is a minimum at the magnetic poles (0 GV) and increases to a maximum of ~17 GV at the magnetic equator. The higher the $R_C$, the higher the energy the cosmic-ray particle must have in order to reach and interact with the atmosphere; otherwise, the cosmic-ray particle is deflected back into space by the Earth's magnetic field (Clem et al., 2004). Cosmic-ray particles interacting with atmospheric molecules undergo nuclear reactions, which in turn create hadronic and electromagnetic showers. The secondary radiation (neutrons, protons, mesons, photons, electrons) produced from these showers propagates in all directions toward the Earth and can be detected at the surface. Cosmogenic neutrons result in the production of several interesting radionuclides present in the environment, such as $^3$H, $^7$Be, $^{10}$Be, $^{14}$C, and $^{36}$Cl (Webber and Higbie, 2003; Webber et al., 2007). The spectrum of secondary cosmogenic neutrons at the Earth's surface covers a wide energy range, from thermal to many GeV. The flux varies with magnetic latitude, elevation, the time in the sun's magnetic activity cycle, and nearby materials, especially ground moisture

(Goldhagen, 2003). There is also a second, albeit much lower flux, source of background neutrons: alpha particles emitted by nuclides in the uranium and thorium decay chains interacting with the nuclei of lighter elements in the Earth's crust (Be, F, Al, Si, etc.) and producing neutrons that escape into the atmosphere (Stenkin et al., 2017; Kuzhevskij et al., 2002).

There are a number of neutron monitors located across the Earth measuring the local intensity of neutrons at different $R_C$ (Hughes and Marsden, 1966; Clem and Dorman, 2000). These neutron monitors consist of banks of moderated (with high-density polyethylene, HDPE) gas proportional counters ($^3$He or $BF_3$) encased in lead. In those monitors, high-energy neutrons create hadronic showers in the lead, which include neutrons at energies that can be efficiently detected by the moderated tubes. Neutron monitors are most sensitive to the highest neutron energies (500 MeV – 100 GeV).

The energy spectrum of cosmic-ray-produced neutrons from ~0.001 eV to >10 GeV has been measured by others in a high-altitude airplane and at several locations on the ground using an extended-range multisphere (Bonner sphere) neutron spectrometer (Goldhagen et al., 2004; Goldhagen et al., 2002; Gordon et al., 2004). The ground measurements were made at five locations in the continental U.S. with values of $R_C$ from 1.6 GV to 4.7 GV and elevations from 14 m to 3450 m (Gordon et al., 2004). The energy resolution of the spectrometer used for those measurements was examined in a paper on the design of an improved spectrometer (Goldhagen, 2011). The neutron spectrum can be plotted as fluence rate per logarithmic energy interval vs. neutron energy on a logarithmic scale (Goldhagen et al., 2004; Goldhagen et al., 2002; Gordon et al., 2004). In that representation, the typical cosmic-ray-produced neutron spectrum has four main regions or features: a thermal neutron peak on the ground (or no thermal peak at altitude); a nearly flat region from epithermal energies to ~0.1 MeV; a broad "nuclear evaporation" peak from ~0.1 MeV to ~10 MeV, centered at 1 MeV – 2 MeV and containing fine structure; and a broad high-energy peak centered near 100 MeV.

The ground measurements of the standard secondary neutron spectrum outlined above were originally made for the microelectronics industry, and the analysis in (Gordon et al., 2004) focused on the high-energy part of the spectrum (>10 MeV), which is responsible for single-event upsets in integrated circuits (JEDEC, 2006) and not significantly affected by variations in local materials. Based on these measurements, an analytical model of the neutron flux cutoff dependence developed by Belov, Struminsky,

Yanke, and Dorman (Clem and Dorman, 2000), and a global map of $R_C$ calculated by (Smart and Shea, 2003), a predictive model for the expected flux of secondary neutrons for any location and elevation on Earth was derived. The parameters input to the model are: latitude and longitude (degrees); solar activity level (%, where 0% is the minimum flux for the active Sun, and 100% is the maximum flux for the quiet Sun); and either the elevation (meters or feet), station pressure (mm-Hg, inches-Hg, or millibar/hPa), or atmospheric depth (g·cm$^{-2}$). The result is a scaling factor for the neutron flux relative to the measured rate in New York City (equal to 1.0) (Neutron Flux Calculation, 2017). The uncertainties in this model are ~20%, with the uncertainty for thermal neutrons being up to a factor of 2. Systematic uncertainties are higher near the equator.

The dominant effect that causes changes to the neutron flux is change in the mass per unit area of air above, called atmospheric depth. For a constant acceleration of gravity, the atmospheric depth is directly proportional to the atmospheric pressure, which varies strongly with elevation and fluctuates as weather changes. Atmospheric depth is best determined from actual barometer readings, but when the barometric pressure is not available, it can be approximately determined from the elevation using a model of the average pressure profile of the lower atmosphere (troposphere): $p = ((44331.514 - z)/11880.516)^{5.255877}$ where $p$ is the average pressure in hPa and $z$ is the elevation in meters (JEDEC, 2006). The result from the predictive model for the relative neutron flux vs. elevation is shown in Fig. 1 (*top*), extending up to 3800 m (e.g., highest peaks in the Rocky Mountains or on the Hawaiian islands). Compared to sea level, the flux changes by a factor of 18 at an elevation of 3800 m for fixed $R_C$ and solar activity level (Neutron Flux Calculation, 2017).

Weather also has an effect on the neutron flux; for example, the dramatic drop in the barometric pressure measured at the neutron monitor in Newark, DE (from 760 mm-Hg to 712 mm-Hg over a 96-h period) during Superstorm Sandy in 2012 correlates with a corresponding 60% increase in rate over that time period (Bartol Research Institute, 2012).

The next largest effect is change in $R_C$, which depends primarily on latitude and slightly on longitude. Fig. 1 (*upper center*) shows the model prediction for the variation of neutron flux with latitude at sea level for a fixed longitude of 90° W relative to the neutron flux at New York City (NYC). Starting at the South Pole (-90°, scaling factor = 1.02), the relative flux reaches a minimum just south of the equator

(-5°, scaling factor = 0.60), and a maximum again near the North Pole (90°, scaling factor = 1.02). The relative flux steadily increases from the equator to the latitude of approximately 45º N. From 46º N to 90º N, the model predicts the relative flux is constant.

Changes in longitude and time in the sun's magnetic cycle show smaller effects on the scaling factor. The variation of the scaling factor with longitude is shown in Fig. 1 (*lower center*). The variation with solar activity is shown in Fig. 1 (*bottom*). The solar modulation term in the model is based on neutron monitor measurements (Clem and Dorman, 2000) made over the 11-year magnetic cycle of the Sun. During solar quiescence, the magnetic influence of the Sun is relaxed, resulting in a greater number of cosmic rays entering the inner heliosphere and thus a higher cosmic-ray flux at Earth. During solar maximum when the Sun is active, marked with sunspots and producing flares and coronal mass ejections, the cosmic-ray flux is reduced as a result of the Sun's magnetic influence on the heliosphere.

Other (lesser) effects on the secondary neutron flux include: the diurnal cosmic-ray variation (Thomas et al., 2017) and transient solar events, namely Forbush decreases (Forbush, 1937), and ground level enhancements (GLEs) (Asvestari et al., 2016). While large sporadic GLEs can cause brief spikes in neutron monitor count rates from strong solar storms, Forbush decreases are typically larger in magnitude and can last for several days.

In this paper, we present the first dedicated campaign to measure secondary neutrons at ground level within close proximity to the magnetic North Pole. We also took data at slightly higher $R_C$ at locations with similar elevations and various ground moisture, and will present comparisons of the results.

2. Experimental Methodology

In June of 2016, Natural Resources Canada (NRCan) conducted a campaign to measure the local neutron background flux at Canadian Forces Station (CFS) Alert (82º N, $R_C$ = 0 GV). Neutron-sensitive instrumentation provided by both the U.S. Naval Research Laboratory (NRL) and NRCan sampled different CFS Alert locations, with varying elevation and ground moisture, in an effort to determine the average count rate and the effects these parameters have on the rate. In November of 2016, a corresponding follow-on campaign was undertaken in southern Canada (~45º N), which allowed us to compare the flux at elevations similar to the measurements at CFS Alert but for slightly higher geomagnetic cutoff rigidity ($R_C$

~1.5 GV). We made measurements using two arrays of $^3$He gas-filled proportional counter tubes, one unmoderated and the other moderated, and plastic scintillators (EJ-299-33) capable of pulse-shape discrimination. Additionally, we monitored background gamma radiation with thallium-doped sodium iodide (NaI:Tl) detectors. A portable weather station was used to monitor pressure at each location. Where feasible, samples were taken of the surface material to measure its moisture content.

NRL provided the unmoderated $^3$He detectors (Radpack-GC) and the EJ-299-33 detectors. NRCan provided the moderated $^3$He detectors (RSN-4), the NaI:Tl gamma detectors, and the weather station. NRCan conducted the measurements at CFS Alert and performed the surface sampling and analysis.

2.1 Detectors

2.1.1 Unmoderated $^3$He detectors (Radpack-GC)

The Radpack-GC, manufactured by (Sensor Technology Engineering, Inc., 2017), consists of seven unmoderated $^3$He gas-filled tubes (ø 5.1 cm × 132.1 cm) at a fill pressure of 2.7 atm (Fig. 2, *top*). The detectors are arranged in two layers; the first layer contains five of the tubes while the second layer contains the other two tubes with a 1.27 cm-thick foam insert between them, abutting the adjacent tubes, for support. The Radpack-GC is contained inside thin nylon housing to minimize moderating material. The proportional counters provide no spectroscopic information; the counts-per-second output is the sum of all seven detectors and is read out via a serial port.

2.1.2 Moderated $^3$He neutron detectors (RSN-4)

The RSN-4, manufactured by (Radiation Solutions Inc., 2017), consists of four $^3$He gas-filled tubes (ø 5.1 cm × 81.3 cm) at a fill pressure of 2.7 atm (Fig. 2, *center*). The tubes are arranged as a 2 × 2 array inside of a thin aluminum box with 1.27 cm-thick HDPE moderator separating each detector and 1.91 cm-thick HDPE moderator enclosing the $^3$He array. The HDPE moderator provides increased sensitivity to incident neutrons in a broad energy range from epithermal to over 1 MeV (see Section 3).

2.1.3 Fast-neutron detectors (EJ-299-33)

We made the fast-neutron measurements with a pair of cylindrical EJ-299-33 scintillation detectors (ø 10.2 cm × 10.2 cm) each coupled to a 7.6-cm photomultiplier tube (PMT) for scintillation light readout. An 80-MHz Bridgeport USB base provided the PMT bias high voltage and signal readout (Bridgeport usbBase). We tuned the EJ-299-33 detectors to have a fast-neutron (recoil-proton) energy range of 700 keV to 15 MeV (proton equivalent energy), corresponding to 80 keV to 6 MeV (electron equivalent energy). We collected data from the Radpack-GC and Bridgeport USB bases with a laptop running custom-written data acquisition (DAQ) software.

2.1.4 Gamma detectors (RSX-1)

Natural gamma radiation was measured using two 10.2 cm x 10.2 cm x 40.6 cm NaI:Tl crystal detectors. The detectors record gamma spectra once per second in 1024 energy bins over the range of 0 keV to 3072 keV.

2.1.5 Location measurement

The Radiation Solutions Inc. (RSI) detection system features a simple global navigation satellite system (GNSS) antenna and receiver, and position and times from the satellite systems are included second-by-second with the radiation measurements in the data record (logged by the RS-700). Additionally, a simple Garmin hand-held GNSS device was used to corroborate the latitude and longitude measurements coming from the RSI instrument. Elevation was then determined by reference of the latitude and longitude locations with the topographic data as represented in Google Earth (CFS Alert, 2017).

2.1.6 Weather Station

At CFS Alert, we used a Vaisala Weather Transmitter (WXT510) to record the following six weather parameters: wind speed and direction, precipitation, atmospheric pressure, temperature and relative humidity. The weather station was used at all locations sampled at CFS Alert. For the campaigns in southern Canada, however, we did not use the weather station; the atmospheric pressure for these locations was determined by the nearest local weather station available.

2.1.7 Surface sampling

Where possible, we performed a sampling of the surface material to determine the concentration of underlying moisture. The upper few centimetres of surface – mainly organic material – was scraped away and not collected. A sample of the underlying surface was then collected using a small shovel to scoop the surface into a plastic Ziploc bag. The sealed Ziploc bag was sealed within a second Ziploc bag so that there could be no moisture loss, and the sample was transported back to Ottawa. At the Geological Survey of Canada laboratory, the samples were subjected to a temperature of 105º C to dry them, and the dry weight was subtracted from the original weight in order to determine the quantity of water. The water content by mass was then formed from the ratio of the water mass to the dry sample mass according to (ASTM D2216 – 10, 2018).

2.2 Data Collection

At CFS Alert, we sampled four locations over bare ground at various elevations to demonstrate the effect that elevation has on rate. The locations and elevations were: sea level (82.5º N, 62.1º W, 0 m), the CFS Alert runway (82.5º N, 62.3º W, 96 m), a transmission station (TX) (82.4º N, 62.5º W, 178 m) and atop a mountain (82.5º N, 64.6º W, 969 m). Images from various site locations are shown in Fig. 3. To minimize shadowing of the ground by the detectors, the effects of small-scale inhomogeneities in the ground, and interference between the detectors, we placed each detector array on plastic sawhorses elevated ~1 m above the ground and separated each unit a few meters from the others. Additionally, we acquired data on a several-meter-thick snow bank to determine the size of the effect on the rate when the measurement is over snow (water) instead of soil or rock (82.5º N, 62.5º W, 141 m; see Fig. 3, *bottom*). For the follow-on campaign in southern Canada, we used the same experimental setup as at CFS Alert excluding the weather station. We acquired data in Ottawa, ON near the NRCan facility (45.4º N, 75.7º W, 80 m) and atop Mont Tremblant in Quebec (46.2º N, 74.6º W, 875 m). Depending on the location during each campaign, we acquired data ranging from a couple of hours to overnight to achieve Gaussian statistics with the thermal and epithermal sensitive detectors. The count rate in these instruments is on the order of a few Hz, yielding $10^4 - 10^5$ total counts, leading to statistical uncertainties of ~0.3% – 1%. The fast-neutron detectors, which count at a rate of ~0.01 Hz, acquired ~2 x $10^2$ total counts per acquisition, resulting in larger errors associated with these measurements. These datasets are summarized in Tables 1 and 2. Lastly, we acquired measurements of the gamma radiation background at each location during the 2016 campaigns.

3. Modelling of thermal neutron detectors

In order to understand the differences in count rate, a comparison of the response functions of the two detectors using $^3$He proportional counters, one with and one without moderator, was performed using the NRL-developed SWORD modeling suite (Novikova et al., 2006). Within the SWORD framework, we used GEANT4 to perform the Monte Carlo calculations. The nuclear data library used by GEANT4 was the G4NDL library, which is primarily derived from the ENDF/B-VII library. For the detector geometries, we included active detection elements (e.g., $^3$He gas, the gas containment vessels, etc.) and the structural elements, such as the cross ties, metal walls, support foam, and integrated moderator. We ignored the detector cabling and electronics in the model and did not include the external supports (the sawhorses).

The response function $R(E)$ is defined such that the detector count rate $c$ in the presence of a neutron flux $\varphi(E)$ can be calculated as $c = \int R(E)\, \varphi(E)\, dE$. To calculate the response for discrete energy values, we define energy bins of equal logarithmic width and calculate the count rate in each bin for an equal fluence in each bin—a "log white" energy spectrum. We define $c_i$, $R_i$, and $\varphi_i$ as the number of counts, the response, and the flux associated with the $i^{th}$ energy bin, respectively. We can then say:

$$c = \sum_i c_i \qquad (1)$$

where

$$c_i = R_i \cdot \varphi_i \qquad (2)$$

Similarly, we can relate the total counts $C_i$ in the $i^{th}$ bin over a given period of time to the associated fluence $\Phi_i$:

$$C_i = R_i \cdot \Phi_i \tag{3}$$

The discrete response is therefore

$$R_i = \frac{C_i}{\Phi_i} \tag{4}$$

To simulate, we shoot a rectangular surface source of $n$ neutrons inward with a constant emission rate per solid angle. The emission surface for each detector was geometrically similar to the outer surface of the detector and enclosed the detector in such a way that, for each face of the detector's outer surface, the corresponding face of the emission surface was located 0.05 mm away. For an isotropic emission, the inward emitted neutrons $n_i$ for the $i^{th}$ energy bin can be related to the fluence $\Phi_i$ and the surface area $A$ of the outer surface of the detector by integrating over the inward solid angle:

$$n_i = \int \left(\frac{\Phi_i}{4\pi}\right) \cdot A \cdot \cos\theta \cdot d\Omega = \frac{\Phi_i \cdot A}{4} \tag{5}$$

where the factor of $\Phi_i/4\pi$ is the fluence per steradian for an isotropic fluence. The total counts $C_i$ are output from the SWORD simulation. We therefore determine $R_i$ as:

$$R_i = \frac{A \cdot C_i}{4 \cdot n_i} \tag{6}$$

Fig. 2 (*bottom*) shows the responses $R_i$ of the RSN-4 and Radpack-GC generated using SWORD. The Radpack-GC shows peak sensitivity to thermal neutrons with some response extending into the epithermal region, while the RSN-4 shows sensitivity over a very broad energy range with most of the response in the epithermal to fast-neutron energy regions.

4. Analysis and Results

In the subsections that follow, we show the results from the thermal and epithermal-to-fast detectors, taken on rock, ground surface, and snow-covered surfaces; fast-neutron measurements for varying elevations and ground moisture content; and gamma-ray measurements taken at all locations.

4.1 Thermal and Epithermal-to-Fast Neutrons

Shown in Fig. 4 are the average neutron count rate vs. atmospheric pressure acquired at CFS Alert, Mont Tremblant and Ottawa. The filled symbol data points are from the moderated $^3$He tubes (RSN-4), and the open symbol data points are from the unmoderated $^3$He tubes (Radpack-GC). The dashed curve shows the predicted relative neutron flux scaling factor (as discussed in Section 2) as a function of atmospheric pressure for the latitude and longitude of CFS Alert, and for a solar activity level of 60% as determined by Thule, Greenland, Neutron Monitor data (Bartol Research Institute, 2016). Note that the relative neutron flux scaling factor is predicted to be nearly identical between CFS Alert and the locations we sampled in the southern Canadian provinces. We normalized the relative neutron flux scaling factor to account for sea-level pressure at CFS Alert, Ottawa, and Mont Tremblant (NOAA, 2016) and scaled by a multiplicative factor such that the curve overlays with the high atmospheric pressure data. Differences in the underlying ground moisture content were not taken into account. In Fig. 4, we find that there is good agreement between the model and the data, except for the point atop the mountain near CFS Alert. On that mountain, both sets of neutron-sensitive instrumentation show count rates higher than expected (unmoderated 20% above the dashed curve, moderated 70% above the dashed curve).

It is possible that part of the discrepancy between the model and data for low atmospheric pressure is due to differences in the ground moisture. The ground-moisture effect is well known, with research dating back more than half a century (Hendrick and Edge, 1966). Calculations from (Hendrick and Edge, 1966) show that the effect of ground water on neutron flux indicates that the ratio of the flux over perfectly dry earth to the flux over ground with ~4% water is roughly a factor of 2. This effect was used by COSMOS (the Cosmic-ray Soil Moisture Observing System) to measure ground moisture with moderated small $^3$He gas proportional counters distributed throughout much of the US (Zreda et al., 2012). Our group at NRL showed the effect of rainfall and soil moisture on background neutron measurements using a large (several square meter) field-deployable array of moderated BF$_3$ and $^3$He gas proportional counters in Florida, US (Hutcheson et al., 2017). We found that variations in the measured neutron counts on the order of 10% were observed when the moisture changed from 100% saturated to 94% saturated. (Stenkin et al., 2017) measured the effect of the seasonal rainfall on thermal neutron count rates indoors high elevation, finding that the ground-moisture effect on the count rate should be ~10%. This effect is not, however, included in the model prediction of the neutron count rate vs. atmospheric pressure or elevation. The ground atop the mountain consists of dry rocks with patches of snow, and running water where the snow has melted, either on the surface or beneath the rock layer. It was not possible to take a surface sample to determine the average ground moisture in this environment (as is shown in the photograph of the ground presented in Fig. 3, *top*). Based on our experience at CFS Alert, we only have uncertain and subjective visual assessments of the ground water content. As indicated in Table 1, measurement of the ground water was possible at three locations – sea level, runway, and TX. Based on our results, we did not observe much effect on the neutron flux compared to the model prediction for the 5% to 7% ground water content at sea level and runway locations versus the 12.5% ground water content at TX. Other possible explanations might be the effect of the coarse array of solid rocks at that location on production and escape of fast neutrons or a different elemental composition of the rocks at that location from the ground at the other locations, affecting the production rate of evaporation neutrons (Zreda et al., 2012).

4.2 Measurements on moisture (snow/ice)

As part of our effort to measure the effects of moisture on the neutron count rate, we acquired both intentional and incidental data on water in the form of snow and ice. First, we took data during the CFS Alert campaign on a several-meter-thick snow bank (Snow). Then, we took data at a similar latitude, longitude and elevation (30 m difference, accounting for a 1% correction), on ground absent of snow (TX). Fig. 5 shows the average count rate from the Radpack-GC and the RSN-4 detectors on snow (Snow), and on ground absent of snow (TX). The Radpack-GC (open symbols) measured virtually no difference (~few percent reduction) in the rate between snow and no snow, while the RSN-4 (filled symbols) measured a much greater reduction (~25%) in the count rate between snow and no snow. For a thick layer of snow, there is not only a lot of moderation thermalizing neutrons, but also less production of neutrons from nuclear evaporation because oxygen is the only available target for spallation/evaporation. Fewer neutrons are available from oxygen compared to other elements in the rock and soil (e.g., Si, Al, Ca, and Fe). Thus, one would expect that the count rate in the moderated detector be significantly lower on the thick snow than on soil, and the rate in the unmoderated detector to be slightly lower. Other campaigns conducted by NRL in the continental US with the Radpack-GC demonstrate results consistent with the CFS Alert data. For instance, during a campaign at NRL in Washington D.C., we acquired data with the Radpack-GC placed on a thin polyethylene table roughly a meter above a surface with no moderator and with a 5-cm-thick moderator present (with the same footprint as the Radpack-GC), resulting in little change in the rate. These results suggest that differences in the moderated/unmoderated detector response lead to differences in the observed count rate with the presence of nearby moderator, supporting the conclusion that underlying ground moisture needs to be taken into account.

As noted in Section 2, during the November 2016 campaign in southern Canada, we wished to acquire data from a similar elevation as the mountaintop at CFS Alert (969 m). We chose Mont Tremblant in the Laurentian mountains of Quebec, a resort in the beginning of the ski (snow-making) season. We conducted an overnight run with our suite of detectors at the summit (875 m). During the acquisition as the night progressed, the resort began the snow-making process for the ski slopes. A shift in the wind pattern led to the instrumentation becoming covered in ~5 cm of snow and ice (Fig. 6, *top*). In this instance, the detector moderator and ground moderator, as opposed to just the ground moderator, had changed. The Radpack-GC DAQ survived the overnight acquisition; however, the RSN-4 DAQ ceased data logging shortly into the acquisition. Fig. 6 (*bottom*) shows the average neutron count rate for the Radpack-GC with a running boxcar average every 300 s for the full acquisition. The dashed curve is the predicted time-dependent relative neutron flux correction based on changes in pressure reported by the Mont Tremblant weather station. The variation in counts for the time duration less than $5.0 \times 10^4$ s tracks with the changing pressure during the night (dashed curve). The sharp (~20%) rise in count rate at between $5.2 \times 10^4$ s and $5.8 \times 10^4$ s is attributed to the accumulation of the snow/ice moderator on the Radpack-GC.

Measurements performed at NRL with moderating material on the Radpack-GC gave similar results (not shown). We found that moderating the top side of the Radpack-GC with 5 cm-thick moderator (HDPE or $H_2O$) produced an increase in the rate by ~20%. Switching the moderator to underneath the Radpack-GC resulted in an increase in rate, but only by ~10%. (Note that the unit was turned over such that the same five-tube side of the Radpack-GC was adjacent to the moderator). This effect may be explained by the underlying surface upon which these data were taken – a building rooftop consisting of mainly concrete. Given that the water-cement ratio – the ratio of the weight of water to the weight of cement used in cement mix – is typically 0.5, the cosmic-ray neutrons would have encountered high hydrogen content underneath the detector, and thus there would be less backscattered neutrons reflected upwards, leading to a discrepancy between moderator on the top and bottom side of the Radpack-GC. The overall lesson from these results was that changing the detector moderator yields a measureable increase in the Radpack-GC rate, as opposed to changing the ground moderator, which has little to no effect on the Radpack-GC rate.

4.3 Fast Neutrons

By using the pulse-shape discrimination method in the EJ-299-33 data set, one can select out fast neutrons from the gamma-ray background. With fast neutrons, we aimed to seek out differences in not only the count rate but also in the spectral distribution. Fig. 7 shows the fast-neutron count rate for all locations sampled at CFS Alert, Ottawa, and Mont Tremblant vs. pressure. The dashed curve is the normalized relative neutron flux correction as a function of pressure. The count rate is on the order of $10^{-2}$ Hz, resulting in larger statistical uncertainties than observed with the thermal and epithermal-to-fast neutron detectors for the same dwell time. Accounting for the statistical uncertainties and possible effects of variation in the elemental composition of the ground, we show that there is general agreement between these data for fast neutrons and the model. For differences in the fast-neutron spectral distribution, we concentrated on three locations at CFS Alert taken at different elevation and ground moisture conditions – 0 m (sea level), 141 m (on snow), and 969 m (on the mountaintop). Fig. 8 (*top*) shows the spectral distribution for each location, normalized to the data from mountaintop. The statistics-limited spectra show no meaningful differences. Based on these results, longer dwell times would be needed to determine if appreciable spectral differences exist for varying elevation. Given that the $R_C$ for CFS Alert and the Ottawa region are similar (0 GV vs. 1.5 GV), we acquired data with longer dwell times at differing elevations near Ottawa (80 m) and atop Mont Tremblant (875 m) to search for elevation-dependent differences in the fast neutron spectral distribution. Fig. 8 (*bottom*) shows no appreciable difference in the fast-neutron spectral distribution as a function of elevation.

4.4 Gamma radiation

Fig. 9 shows the average gamma-ray count rate for the sum of the pair of NaI:Tl detectors over their full energy range (0 keV – 3072 keV). These measurements encapsulate not only differences in location, but also variation in the underlying surface, namely: rock, soil, and snow. CFS Alert measurements were done on rock and soil at varying elevations; the measurement atop Mont Tremblant was done on soil. The differences in the rate at the CFS Alert and Mont Tremblant locations are attributed to areas with higher concentration of $^{40}$K (1461 keV gamma rays) and $^{208}$Tl (2614 keV gamma rays), as observed in the spectroscopic data. In Ottawa, the measurement on soil shows a reduction in counts by a factor of ~2. Further reduction (by a factor of ~4) was observed in the measurement done on snow at CFS Alert.

5. Conclusions

We have presented the results from measurements taken with neutron (thermal, epithermal-to-fast, and fast) and gamma-ray sensitive detectors near the geomagnetic North Pole and, for comparison, in the southern part of the Canadian provinces of Ontario and Quebec. The main purpose of these measurements was to acquire data near the geomagnetic North Pole to compare with an analytic predictive flux model with its coefficient for atmospheric-depth (atmospheric-pressure) dependence derived from measurements conducted at higher $R_C$. For the neutron detectors with thermal and epithermal-to-fast sensitivity, we find good agreement between these data and the model near the geomagnetic North Pole and in southern Canada, both of which have similar $R_C$ – affirming our initial hypothesis. There is, however, a discrepancy between measurements made atop a mountain at CFS Alert, resulting in a higher rate (20% for unmoderated, 70% for moderated) than the model predicts. This discrepancy might have been caused by differences in the ground moisture content atop the mountain compared to the other locations sampled, or differences in other features of the surface at that location, such as elemental composition, surface roughness, or local natural radioactivity. The data presented show the first measurements of the secondary cosmic-ray neutron background, ranging from thermal to fast, from the Earth's surface near the geomagnetic North Pole. Ground-truth measurements are important and needed for validating the predictive flux model that can be applied to any location on Earth.

Acknowledgements

For NRL, this work was sponsored by the Chief of Naval Research (CNR). For NRCan, the NRCan Contribution number: 20170337. The authors would like to thank Dr. Paul Goldhagen from the U. S. Department of Homeland Security National Urban Security Technology Laboratory for the helpful discussions regarding the data procurement procedure, Dr. Lee Mitchell for assisting with measurements at NRL, Dr. David Waller and Mr. Ian Watson from Defence Research and Development Canada and Mr. Shawn Usman from the National Geospatial-Intelligence Agency for assisting in the arrangement of this campaign. We acknowledge the University of Delaware Bartol Research Institute for providing the neutron monitor data that was cited in this manuscript. Additionally we wish to thank the Canadian Armed Forces who went out of their way to provide personnel safety, personnel transportation, and logistical support at CFS Alert. Environment Canada also provided assistance at CFS Alert. Lastly, we gratefully acknowledge the assistance of Tremblant ski resort, Mont Tremblant, Quebec.

**Tables**

Table 1: Description of data sets acquired at CFS Alert, Nunavut, Canada.

| Dataset Location | Latitude/Longitude | Elevation (meters) | Ground Moisture (water content, %) | Notes |
|---|---|---|---|---|
| Sea Level | 82.5º N, 62.1º W | 0 | 7.0 | Sand and gravel |
| Runway | 82.5º N, 62.3º W | 96 | 5.3 | Reworked gravel |
| Snow | 82.5º N, 62.5º W | 141 | (100) | Snow. See Fig. 3 (*bottom*). |
| Transmission Station (TX) | 82.4º N, 62.5º W | 178 | 12.5 | Till. See Fig. 3 (*center*). |
| Mountaintop | 82.5º N, 64.6º W | 969 | - | Rock with interspersed snow and running water. Surface sample not possible. See Fig. 3 (*top*). |

Table 2: Description of data sets acquired in southern Canada. Note that no weather station data or surface samples were acquired for this part of the campaign.

| Dataset Location | Latitude/Longitude | Elevation (meters) |
| --- | --- | --- |
| Ottawa, Ontario | 45.4º N, 75.4º W | 80 |
| Mont Tremblant, Quebec | 46.4º N, 74.3º W | 875 |

Figures

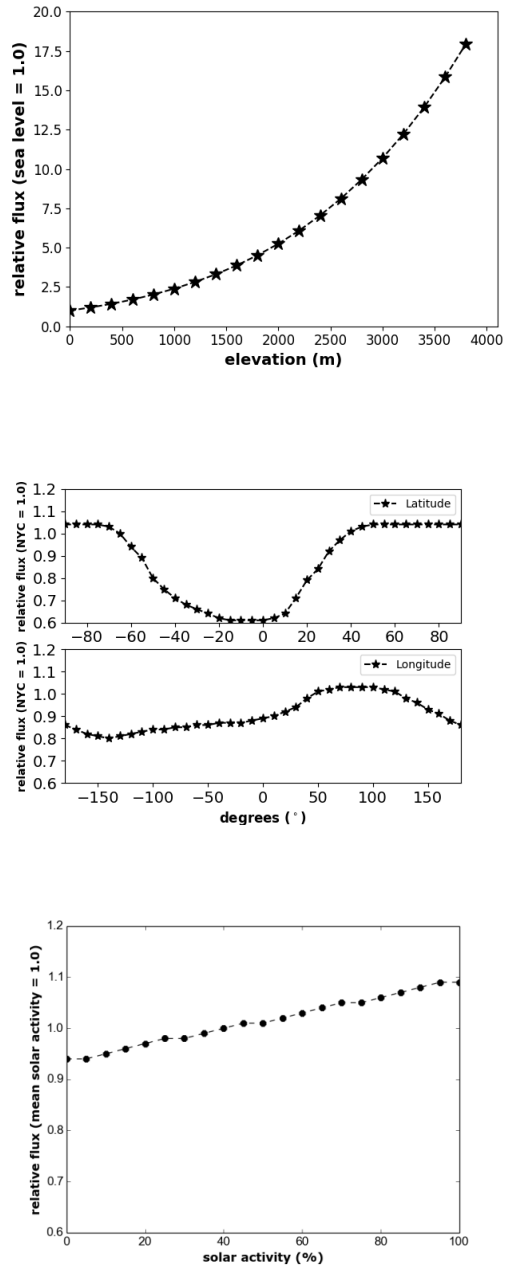

Fig. 1. The effect on the relative neutron flux scaling factor for varying: elevation with location such that $R_C$ is equal to that of NYC (*top*); latitude (at sea level for a fixed longitude of 90º W) and longitude (at sea level for a fixed latitude of 45º N) (*center*); and, solar activity (at sea level for a fixed latitude 45º N and longitude 90º W) (*bottom*). These data were obtained from the predictive model discussed in the introduction and presented in (Neutron Flux Calculation, 2017).

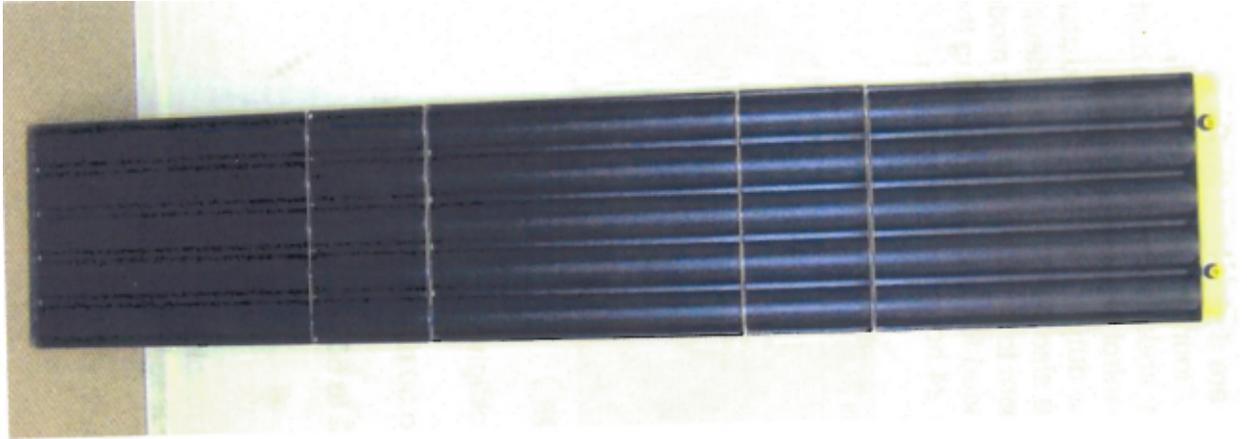
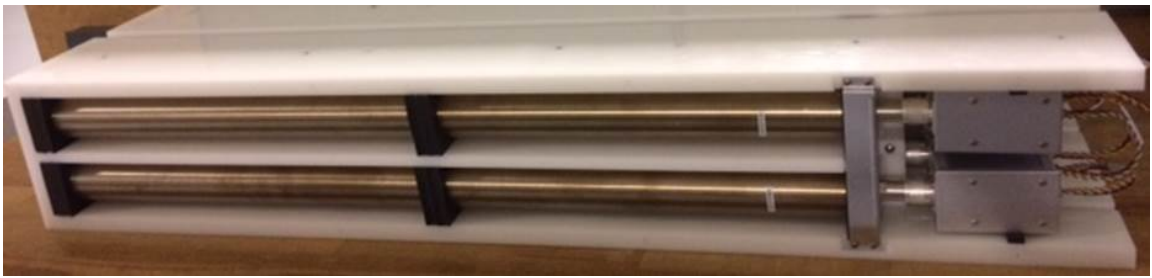
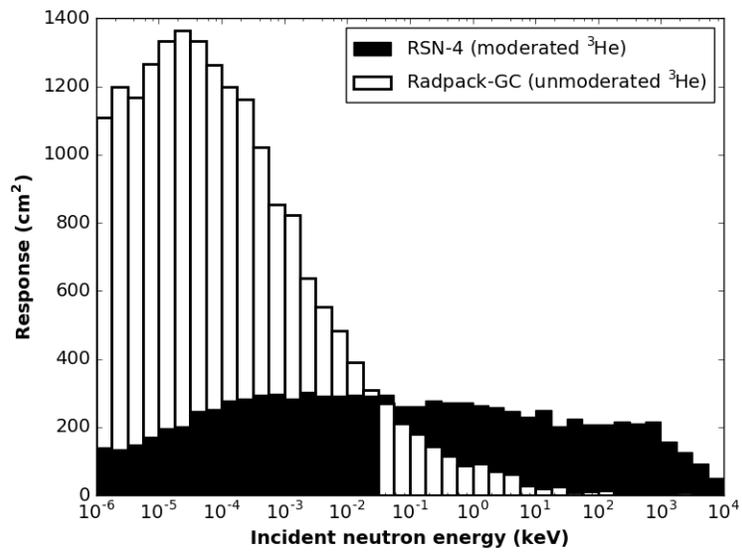

Fig. 2. The NRL Radpack-GC (*top*), the NRCan RSN-4 (*center*), and the simulated response for each detector (*bottom*).

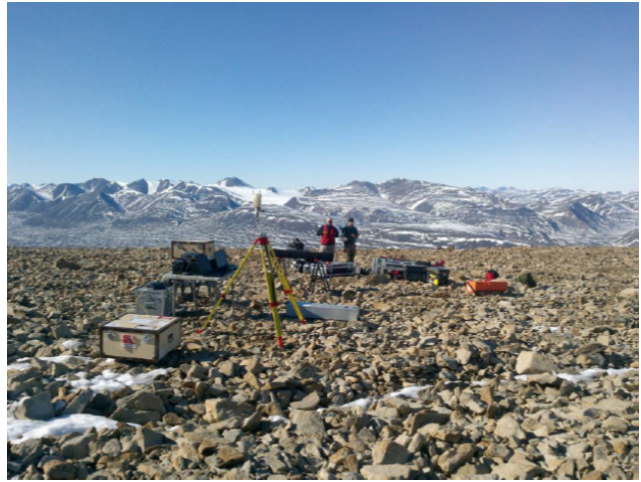
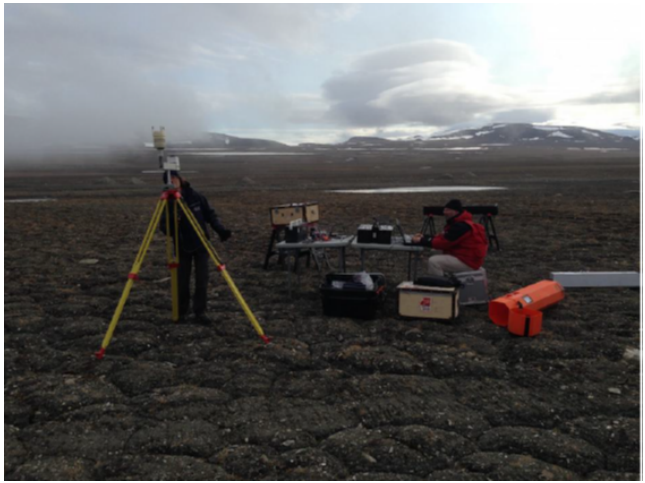
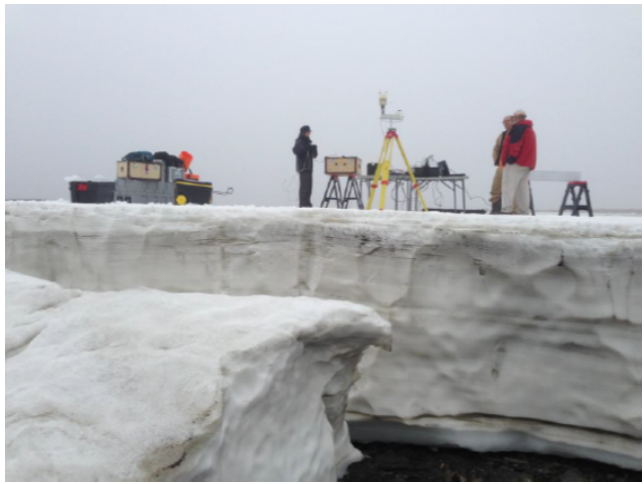

Fig. 3. Images from sampled locations around CFS Alert. On a mountaintop (*top*); near the transmission station (TX) (*center*); on a snow bank (*bottom*).

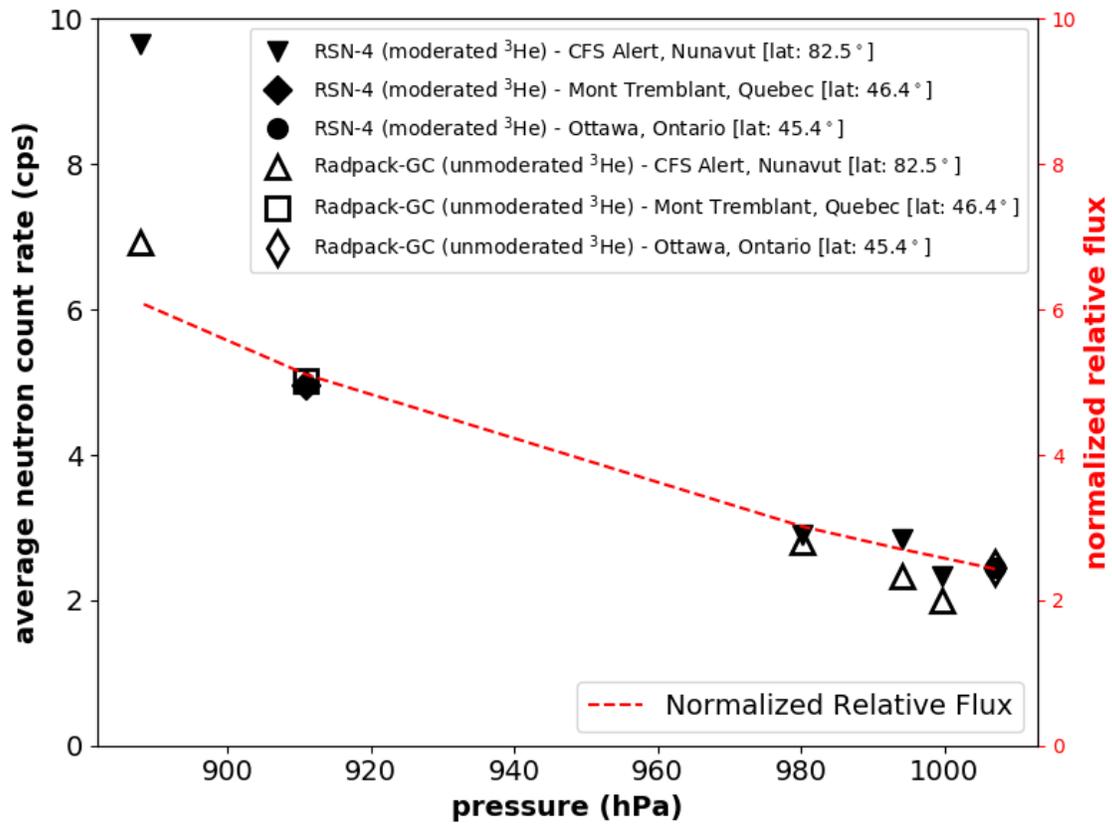

Fig. 4. Average neutron count rate vs. pressure. Dashed curve shows the relative normalized flux derived from the (Neutron Flux Calculation, 2017). The RSN-4 and Radpack-GC data sets are normalized according to the measurements made in Ottawa, Ontario. The symbols representing these data at this location (filled circle and open diamond) are overlaid on top of each other. Note that the error bars for the average count rate for the Radpack-GC and RSN-4 are smaller than the symbols used to represent these quantities.

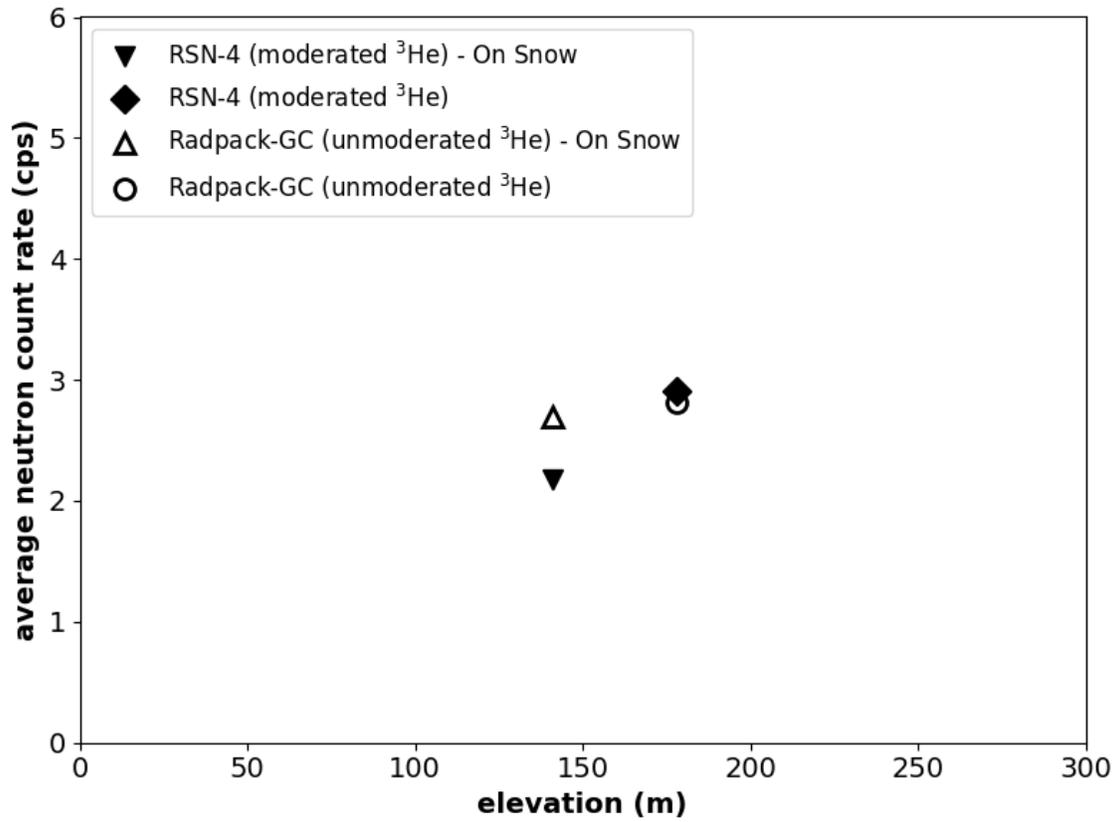

Fig. 5. Radpack-GC (unmoderated $^3$He) and RSN-4 (moderated $^3$He) measurements made above a snow covered surface and a nearby surface absent of snow cover at CFS Alert. Note that the error bars for the average count rate for the Radpack-GC and RSN-4 are smaller than the symbols used to represent these quantities.

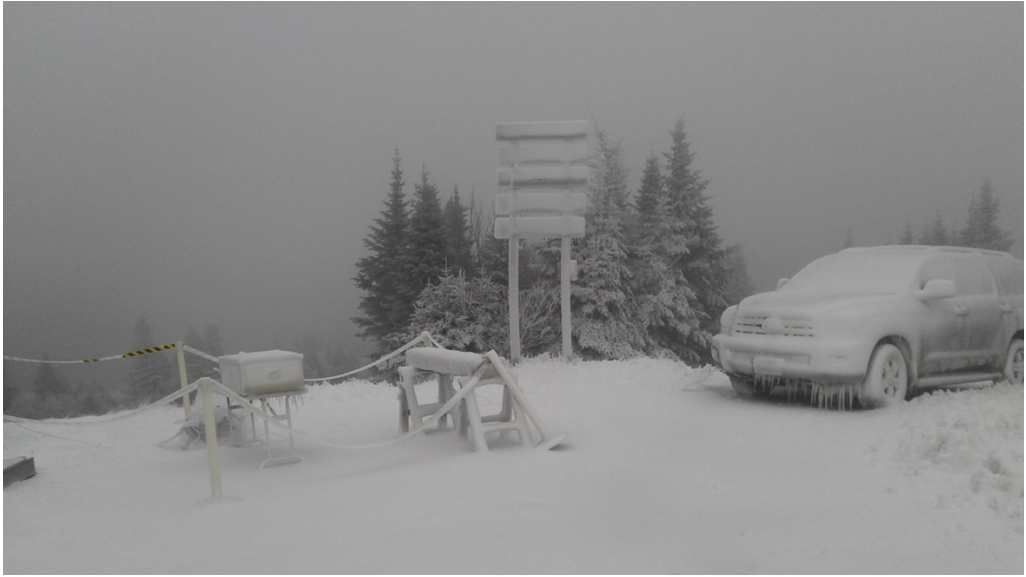

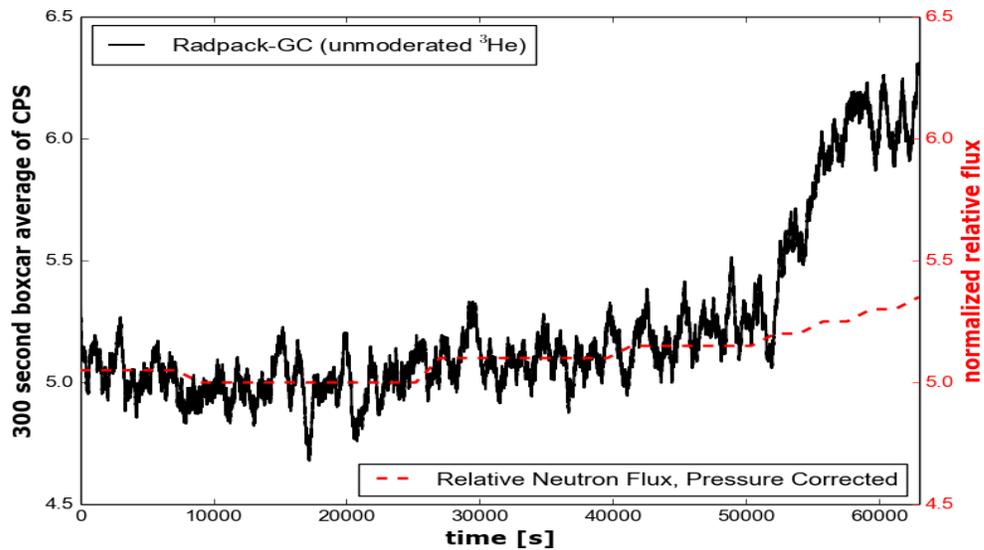

Fig. 6. Neutron-sensitive equipment covered in snow/ice atop Mont Tremblant (*top*). The Radpack-GC (unmoderated $^3$He) rate shown with the pressure-corrected relative neutron flux (dashed curve) (Neutron Flux Calculation, 2017). The sharp increase in counts at ~5.2 x $10^4$ s corresponds to the start of snow/ice covering the equipment (*bottom*).

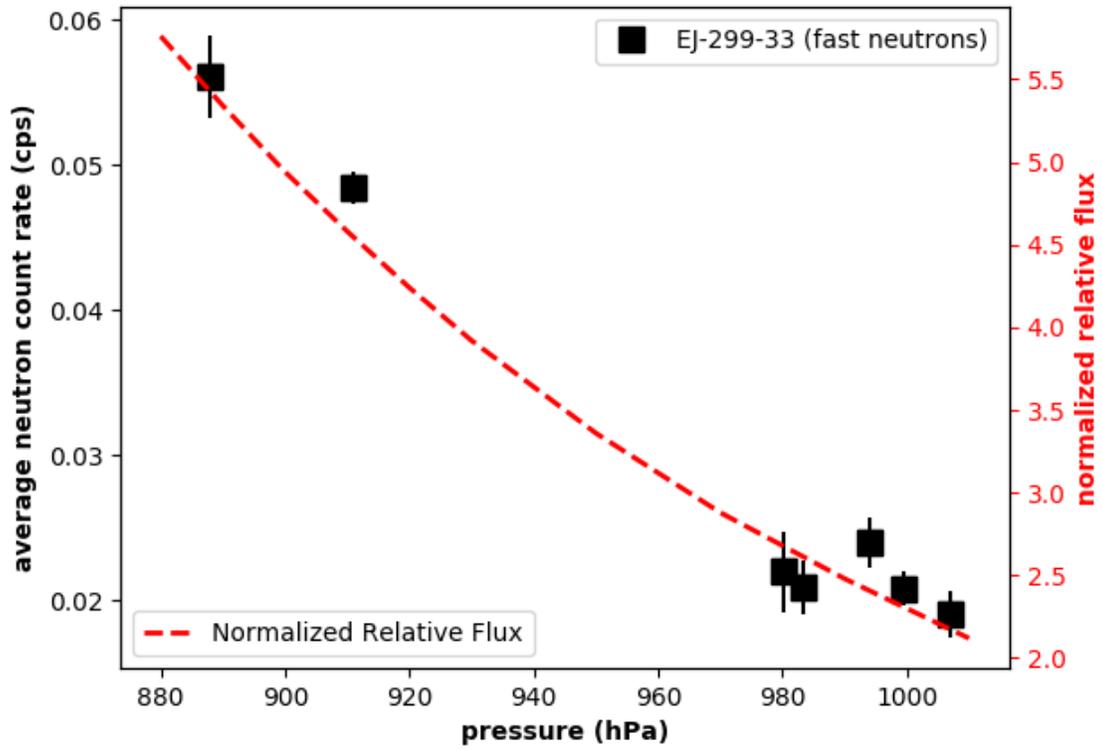

Fig. 7. Average fast-neutron count rate vs. pressure for locations at CFS Alert, Mont Tremblant, QC, and Ottawa, ON. Dashed curve shows the relative normalized flux derived from (Neutron Flux Calculation, 2017).

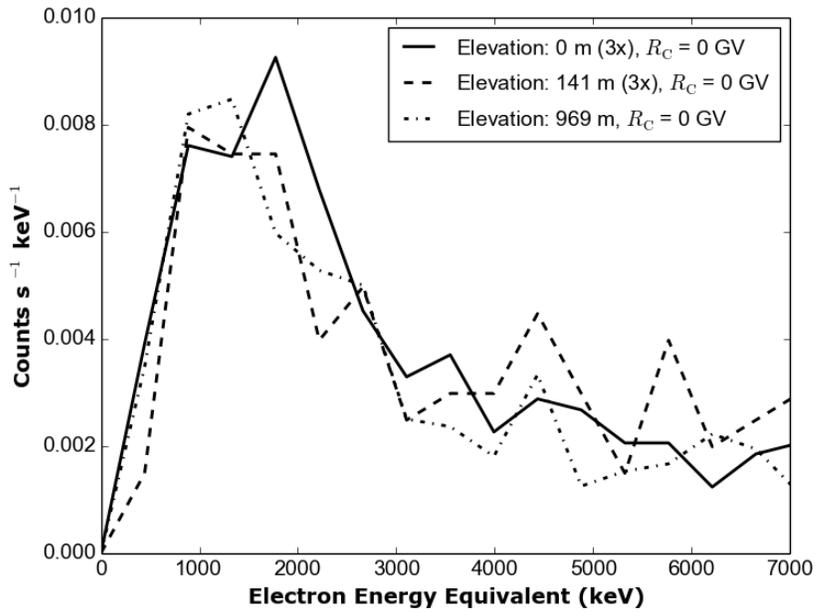

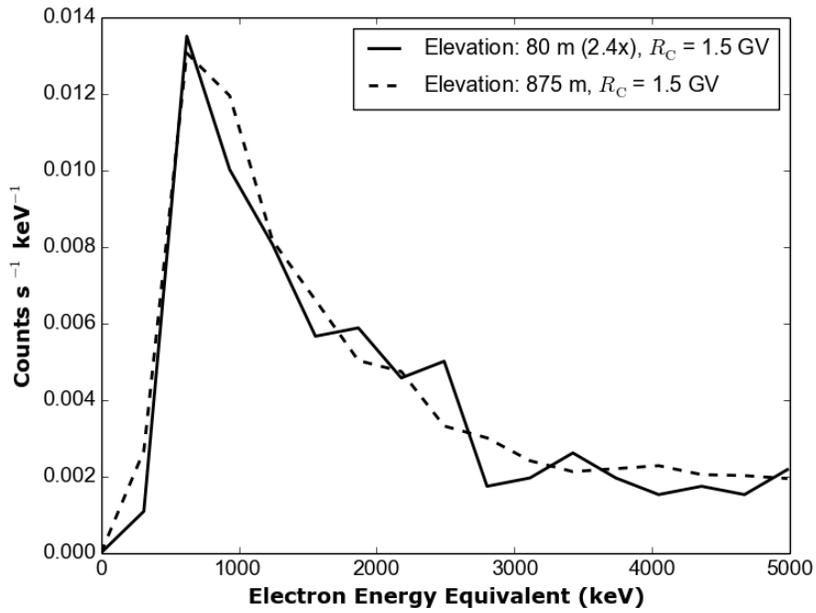

Fig. 8. Normalized fast-neutron spectral distributions acquired by the EJ-299-33 detectors at: CFS Alert at sea level (0 m), on snow (141 m), and on a mountaintop (969 m) (*top*); Ottawa, ON (80 m) and Mont Tremblant, QC (875 m) (*bottom*).

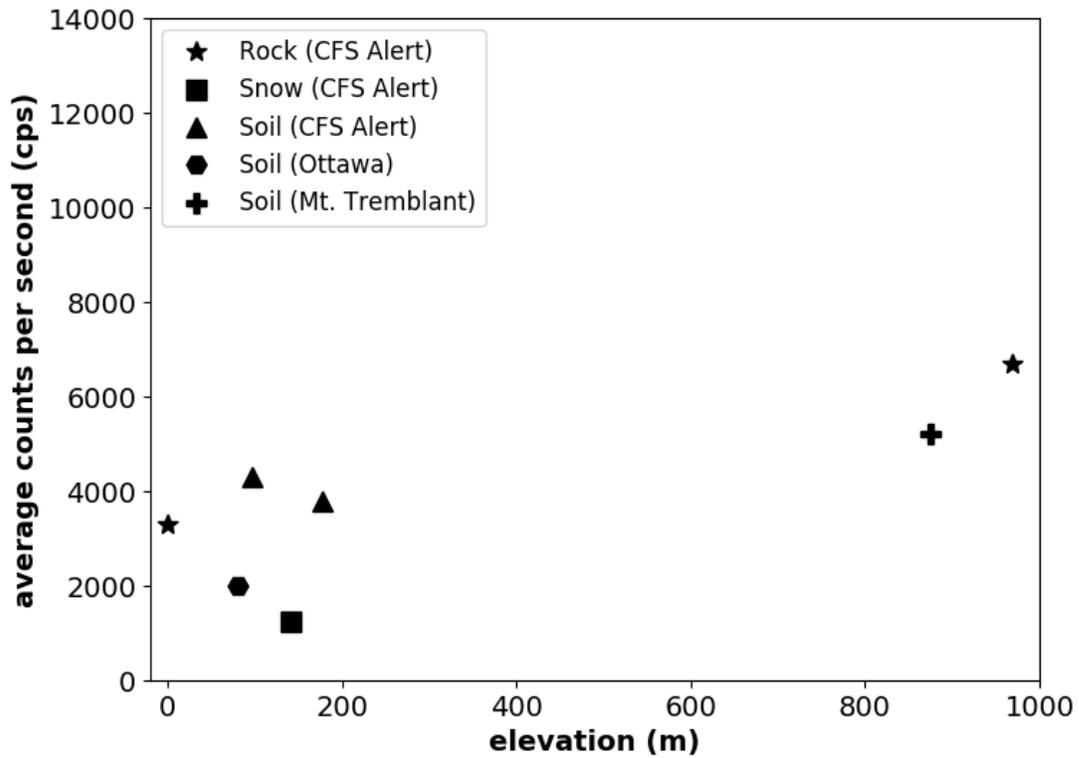

Fig. 9. Average gamma-ray count rate vs. elevation for locations sampled at CFS Alert, Ottawa, ON, and Mont Tremblant, QB with varying underlying surface content. Measurements were made with a pair of 10.2 cm x 10.2 cm x 40.6 cm NaI:Tl crystal detectors. Note that the error bars for the average gamma-ray count rate are smaller than the symbols used to represent these quantities.